# Size-Independent Quantification of Ligand Binding Site Depth in Receptor Proteins*

**Srujana Cheguri[1] and Vicente M. Reyes, Ph.D.[2], [3]**

(*, M. S. Thesis, part 1; [1], M. S. student; [2], thesis advisor)
[3], E-mail: vmrsbi.RIT.biology@gmail.com

Submitted in partial fulfillment of the requirements for the **Master of Science** degree in Bioinformatics at the Rochester Institute of Technology

**Srujana Cheguri**

Dept. of Biological Sciences, School of Life Sciences
Rochester Institute of Technology
One Lomb Memorial Drive, Rochester, NY 14623

November 2011

11-1-2011

# Part 1: Size-independent quantification of ligand binding site depth in receptor proteins. Part 2: Representing rod-shaped protein 3d structures in cylindrical coordinates

Srujana Cheguri

# *Part 1: Size-Independent Quantification of Ligand Binding Site Depth in Receptor Proteins*

# *Part 2: Representing Rod-Shaped Protein 3D Structures in Cylindrical Coordinates*

Srujana Cheguri

**Approved:** ______________________________

**Vicente Reyes, Ph.D.**
*Thesis Advisor*

______________________________

**Gary Skuse, Ph.D.**
*Committee Member*

______________________________

**Paul Craig, Ph.D.**
*Committee Member*

Submitted in partial fulfillment of the requirements for the degree of Master of Science

Department of Bioinformatics
College of Science
Rochester Institute of Technology
Rochester, NY

November, 2011

***This thesis is dedicated to my beloved family, to my parents for their never-ending encouragement, love and confidence in me and to my brother for his motivation and guidance.***

**ACKNOWLEDGEMENTS**

I feel it as a unique privilege, combined with immense happiness, to acknowledge the contributions and support of all the wonderful people who have been responsible for the completion of my master’s degree. The two and half years of graduate study at RIT has taught me that creative instinct, excellent fellowship and perceptiveness are the very essence of science. They not only impart knowledge but also place emphasis on the overall development of an individual. I am extremely appreciative of RIT, especially the Department of Biological Sciences (Bioinformatics Option) in this regard. I owe it to my mentors at RIT to what I am today.

I would like to express my deepest gratitude to my thesis advisor, Dr. Vicente Reyes who continually encouraged and guided me during the course of my thesis work. I would also like to express my appreciation to my committee members Dr. Gary Skuse and Dr. Paul Craig for their valuable guidance, timely help and support. A special thanks to Nicoletta Bruno Collins for all the academic formalities that was needed to be done.

I would like to thank all professors, mentors, family and friends who have helped me scale heights and achieve this prestigious degree at RIT. Finally, I would like to thank RIT for giving me an opportunity to be a part of this family.

## LIST OF ABBREVIATIONS

| | |
|---|---|
| SPi | Secant Plane Index |
| TSi | Tangent Sphere Index |
| SPM | Secant Plane Method |
| TSM | Tangent Sphere Method |
| GC | Global Centroid |
| LC | Local Centroid |
| LBS | Ligand Binding Site |
| 3D | Three Dimensional |
| RSP | Rod Shaped Protein |
| PDB | Protein Data Bank |
| $\alpha$ | Alpha |
| $\beta$ | Beta |
| HTML | Hyper Text Markup Language |
| PHP | Hypertext Preprocessor |
| CSS | Cascading Style Sheets |
| GUI | Graphical User Interface |
| KB | Kilo Bytes |
| | |

## ABSTRACT FOR PART 1:

We have developed a web server that implements the two complementary methods to quantify the depth of ligand and/or ligand binding site (LBS) in a protein-ligand complex. The two methods are the 'secant plane' (SP) and the 'tangent sphere' (TS) methods. In the SP and TS methods, the protein molecular centroid (global centroid, GC), and the LBS centroid (local centroid, LC) are first determined. The SP is defined as the plane passing through the LBS centroid and normal to the line passing through the LC and the protein molecular centroid. The "exterior side" of the SP is the side opposite GC. The TS is defined as the sphere with center at GC and tangent to the SP at LC. The percentage of protein atoms (a.) inside the TS (TSi) and (b.) on the exterior side of the SP (SPi), are two complementary measures of ligand or LBS depth. The SPi is directly proportional to LBS depth while the TSi is inversely proportional to LBS depth. We tested the SP and TS methods using a test set of 67 well characterized protein-ligand structures (Laskowski, et al. 1996), as well as the theoretical case of an artificial protein in the form of a cubic lattice grid of points in the overall shape of a sphere and in which LBS of any depth can be specified. Results from both the SP and TS methods agree very well with reported data (Laskowski, et al. 1996), and results from the theoretical case further confirm that both methods are suitable measures of ligand burial or LBS depth. There are two modes by which one can utilize our web server. In the first mode we term the 'ligand mode', the user inputs the PDB structure coordinates of the protein as well as those of its ligand (one ligand at a time if there is more than one). The second mode, the 'LBS mode', is the same as the first except that the ligand coordinates are assumed to be unavailable; hence the user inputs what s/he believes to be the coordinates of the LBS amino acid residues. In both cases, the web server outputs the SP and TS indices. LBS depth is an important parameter as it is usually directly related to the amount of conformational change a protein undergoes upon ligand binding, and ability to quantify it could allow meaningful comparison of protein flexibility and dynamics. The URL of our web server is http://tortellini.bioinformatics.rit.edu/sxc6274/thesis1.php

## LIST OF FIGURES

## LIST OF TABLES

# Table of Contents

# *Part 1: Size-Independent Quantification of Ligand Binding Site Depth in Receptor Proteins*

# Chapter 1

## *1. Introduction*

How ligands bind their cognate receptor proteins is an important question in biology. Most proteins are flexible, and the ligand-binding event induces a conformational change in both ligand and protein leading to the more stable structure of the protein-ligand complex. The amount of conformational change a protein undergoes varies and is usually directly related to the depth of the ligand-binding site. Therefore, it is very important to be able to determine the ligand binding sites and their depths, as this information provides insight into protein dynamics and flexibility.

### 1.1 Ligand Binding Sites

Although the main objective of this study is not the prediction of ligand binding sites, but instead the quantitative determination of the depth (degree of burial) of ligand binding sites, we present here a few methods for predicting ligand binding sites, as it is a related concept (Alasdair & Richard, 2005). In order to study the protein ligand binding sites, their prediction is necessary. There are mainly two types of methods that are used to predict the ligand binding sites in proteins. They are geometric and non-geometric methods.

### Geometric Methods

Geometric methods take the geometry of the protein into consideration. Generally Geometric based methods are most widely used to detect the ligand binding sites on the proteins. Some of the geometric based methods are ligsite (Hendlich, Rippmann & Barnickel, 1997), pass (Brady & Stouten, 2000), travel depth (Coleman & Sharp, 2006), pocket-picker (Martin, Ewgenij & Gisbert, 2007), surfnet (Laskowski, 1995) and pocket-finder (Alasdair & Richard, 2004). These methods identify the ligand binding sites and also compare different ligand binding sites. But each of them has its own shortcomings.

### Non-Geometric Methods

Non-geometric methods take into account of the interaction energy between protein and the probe and evolutionary information of the protein into consideration. One of the important non-geometric based approaches developed was Q-Site Finder. The success rate for Q-Site Finder is very high when compared to all the geometric based approaches. In ninety percent of the proteins tested in Q-Site Finder there is more than one successful prediction in the top three binding sites. It is one of the important non-geometric methods. It takes evolutionary information of the proteins into consideration and is proved to be more accurate when compared to the other geometric based methods (Alasdair & Richard, 2005).

However, in addition to the prediction of ligand binding sites the determination of the ligand binding site depth in a quantitative way is also necessary for a full understanding of protein-ligand interactions.

## 1.2 Ligand binding site depth

There are limited resources that can quantitatively measure the depth of the ligand binding sites. Some of the geometric based methods such as ligsite (Hendlich, Rippmann & Barnickel, 1997), surfnet (Laskowski, 1995), pocket-finder (Alasdair & Richard, 2004) estimates the depths of the pockets present on the protein surface. They are discussed below.

**Ligsite:** A program that automatically detects pockets on the surface of proteins by binding hydrophobic probes to the proteins in a time-efficient manner. It is used for comparative studies of the proteins for the large set of proteins. It does not give any quantitative results of the protein (Hendlich, Rippmann & Barnickel, 1997).

**Surfnet:** This program generates surfaces and void regions. The program detects the gap regions that are present in between the protein and it isolates the protein from the gap region. The binding site is predicted to be in the largest gap region (Laskowski, 1995).

**Pocket-Finder:** This program is based on ligsite, which detects the ligand binding sites. Pocket Finder predicts the volume of the binding sites, but does not give depth of the binding site (Alasdair & Richard, 2004).

**Mathematical Model:** Depth of the LBS can be determined using Mathematical model. The mathematical model uses accessible radius function theory in spatial particle system as the first step. The second step is applying hierarchical analysis on the peptide chains and proteins and then determining the depth using mathematical function known as depth function algorithm. The equation for the depth of a point 'a' in a set 'A' in the algorithm is defined as:

$$S_A(a) = \min\{n_1(\pi,a): \pi \varepsilon \pi`(a) \}$$

where $\pi`(a)$ is a plane that contains point a. Finally a depth database can be built for all the proteins in the PDB. (Shen & Tuszynski, 2008).

In summary the present methods compare the sizes of the pockets detected in a protein or in proteins of same size and detect the void volumes present in a protein.

# Chapter 2

## *1. Statement of the Problem*

Currently available methods compare LBS in proteins of equal size. It is very important in structural biology studies to determine the ligand binding site depths and quantitatively compare the LBS in proteins of different sizes instead of simply detecting and counting the ligand binding sites.

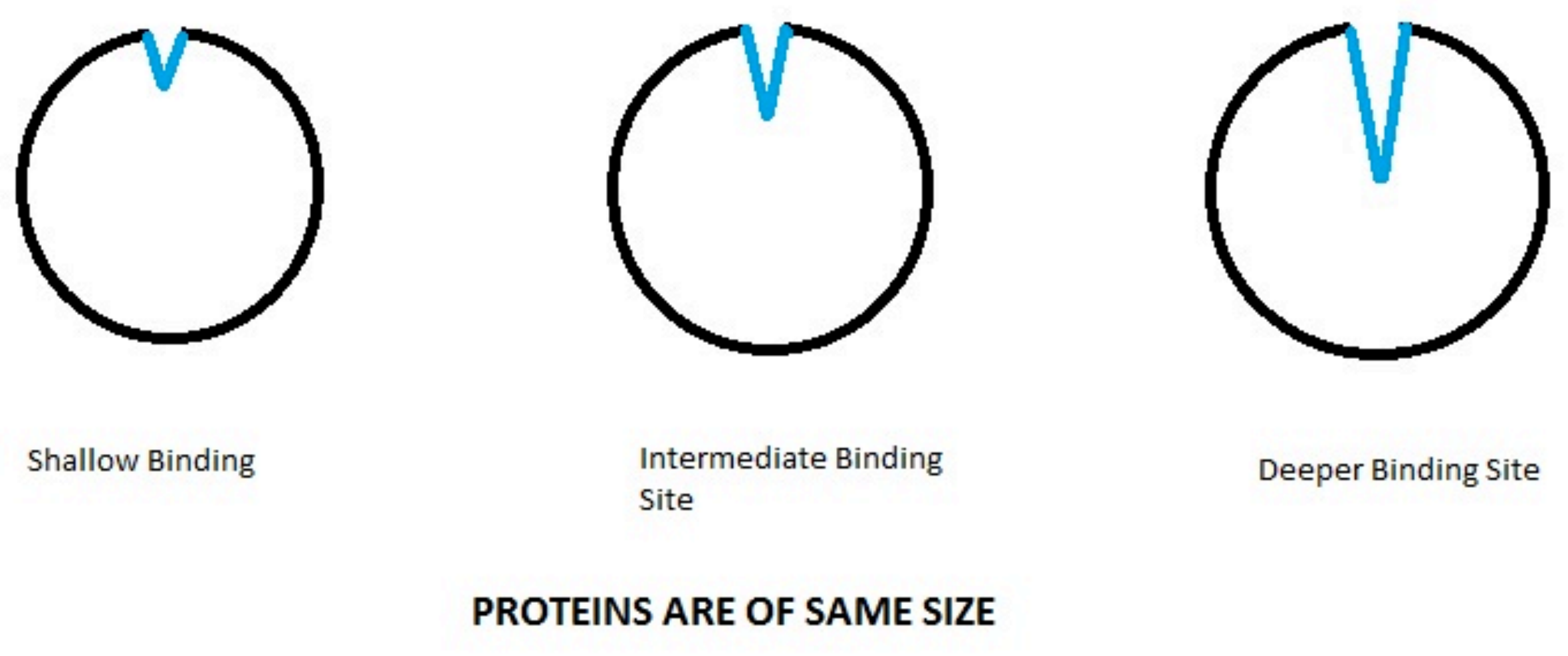

**Figure 1: Proteins are of same size**

In Figure 1 all the three proteins are of same size in which the first protein has shallow binding site, second protein has intermediate depth binding site and the third protein has deepest binding site. As the proteins are of same size we are able to readily compare the ligand binding site depths. The amount of conformational change a protein undergoes is thought to be directly proportional to the depth of the ligand-binding site. We may conclude that the first protein undergoes little conformational change, second protein undergoes intermediate amount of conformational change and the third protein undergoes higher conformational change upon binding their cognate ligands.

But in more of the case in practice, we have proteins of different sizes and their ligand binding site depths have to be compared. For example in Figure 2 the two proteins are of different size but they might have equal ligand binding site depths or volume in absolute terms.

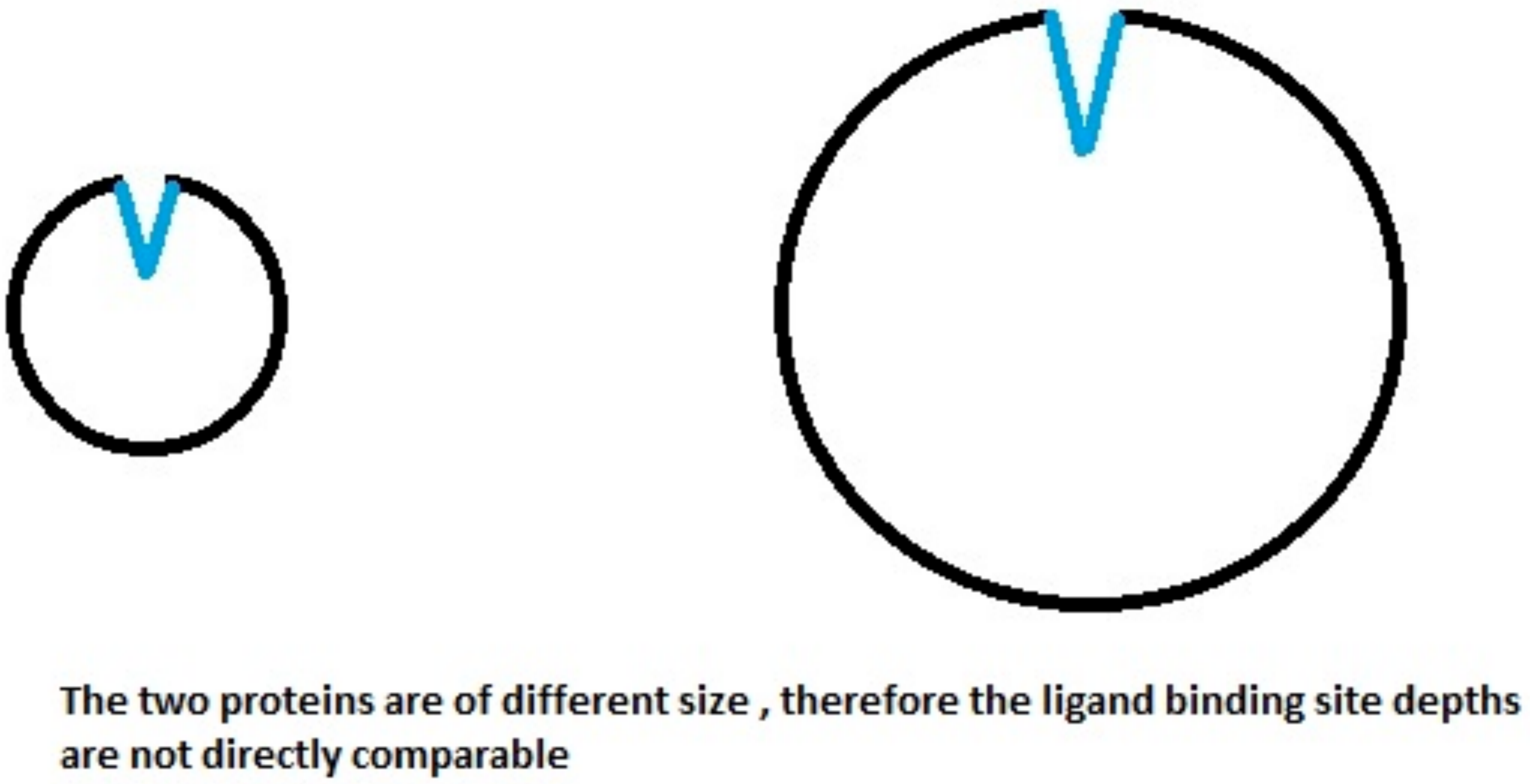

**Figure 2: Proteins are of different sizes**

The aim of this work is to device a quantitative metric for comparing LBS depth (or burial) that takes into account the volume of the protein so that the comparison of LBS depth (or burial) is meaningful.

## Chapter 3

## *1. Methods*

We have developed the Secant Plane Method and Tangent Sphere Method (SPM and TSM, respectively) to quantitatively determine the binding site depths in proteins. To the best of our knowledge, this work is the first quantitative and comparative measure of ligand binding site depth in proteins. SPM and TSM are explained in Figure 3. Here for the Secant Plane and Tangent Sphere Methods we consider the following two points.

a) Protein Centroid: It is the geometric center of the protein, found by analysis of the x, y and z coordinates. It is also known as the Global Centroid.

b) Local Centroid: The centroid of the bound ligand, or a few amino acid residues in the LBS.

### 1.1 Secant Plane Method

We define that the Secant Plane passes through the Local Centroid and is normal to the line passing through the local centroid and the global centroid. The Secant Plane index (SPi) is defined as the percentage of the protein atoms on the exterior side of the Secant Plane

$$\mathbf{SPi} = \frac{\text{Number of protein atoms on the external side of the Secant Plane}}{\text{Total number of atoms}} * 100$$

## 1.2 Tangent Sphere Method

We define tangent sphere as the sphere at the center of the protein and it is tangent to the secant plane and its radius is equal to the distance between global centroid and local centroid. The Tangent Sphere Index (TSi) is defined as the percentage of protein atoms inside the tangent sphere

$$TSi = \frac{\text{Number of protein atoms inside the Tangent Sphere}}{\text{Total number of Atoms}} * 100$$

Depth of LBS burial is directly proportional to the SPi (as shown in the figure 3), the deeper the LBS the higher the SPi. On the other hand, depth of the LBS burial is inversely proportional to the TSi, the deeper the LBS the lower the TSi.

Programs: TSM and SPM are written in Fortran 77 and 90 to calculate the SPi and TSi.

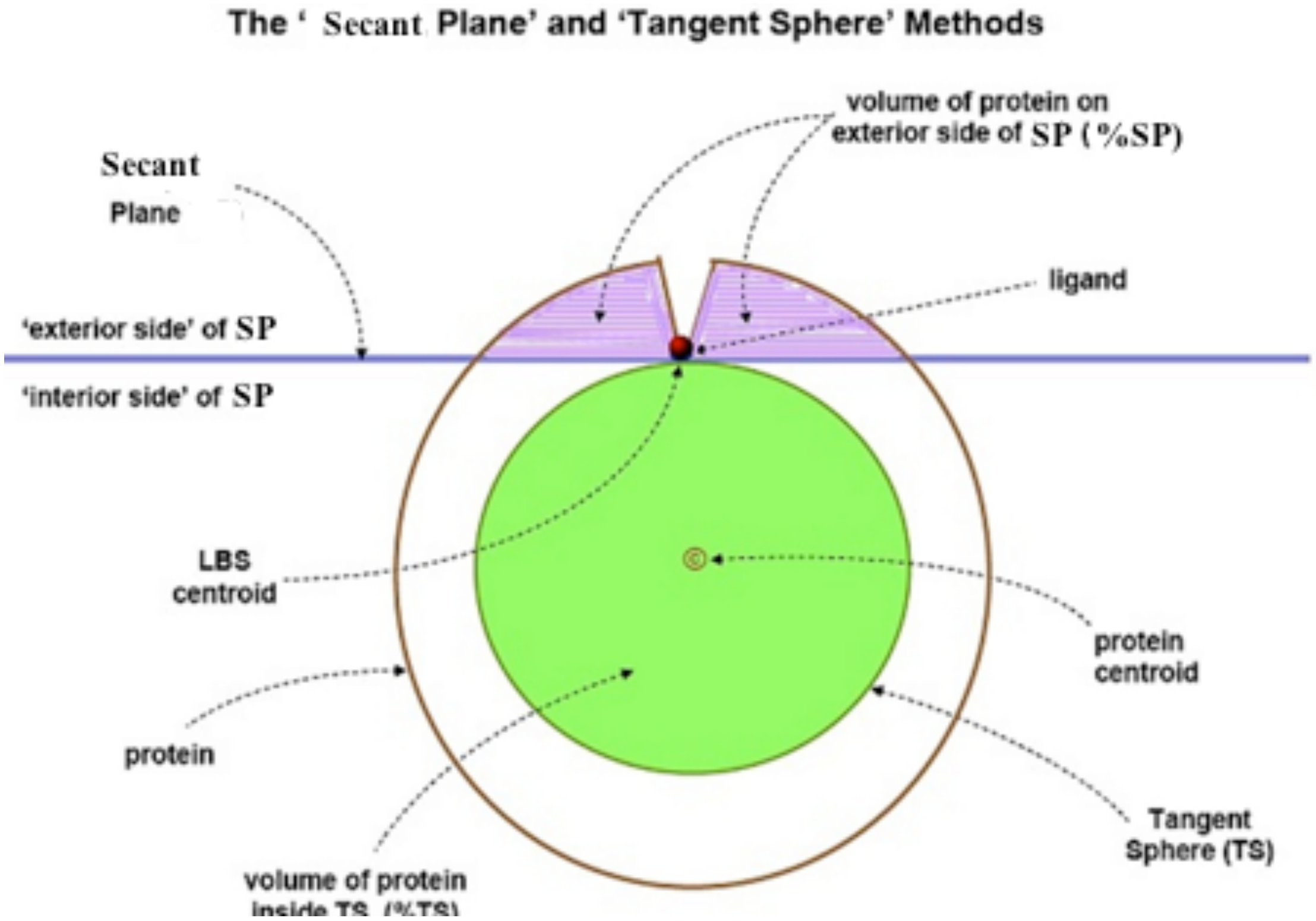

**Figure 3: Secant Plane and Tangent Sphere Methods**

Therefore, the SP and TS methods are complementary to each other. However, they are not redundant, as one cannot be calculated from the other. This is due to the fact that proteins are irregular in shape instead of being in perfect sphere. The equation of the Secant Plane is given by **Ax+By+Cz+D=0** where A, B, C, D are constants and x, y, z are the coordinates of the points that lie on the plane. The points that lie on the exterior side of the Secant Plane satisfy this equation **Ax+By+Cz+D<0** and the points on the other side (interior) of the Secant Plane satisfies this equation **Ax+By+Cz+D>0.**

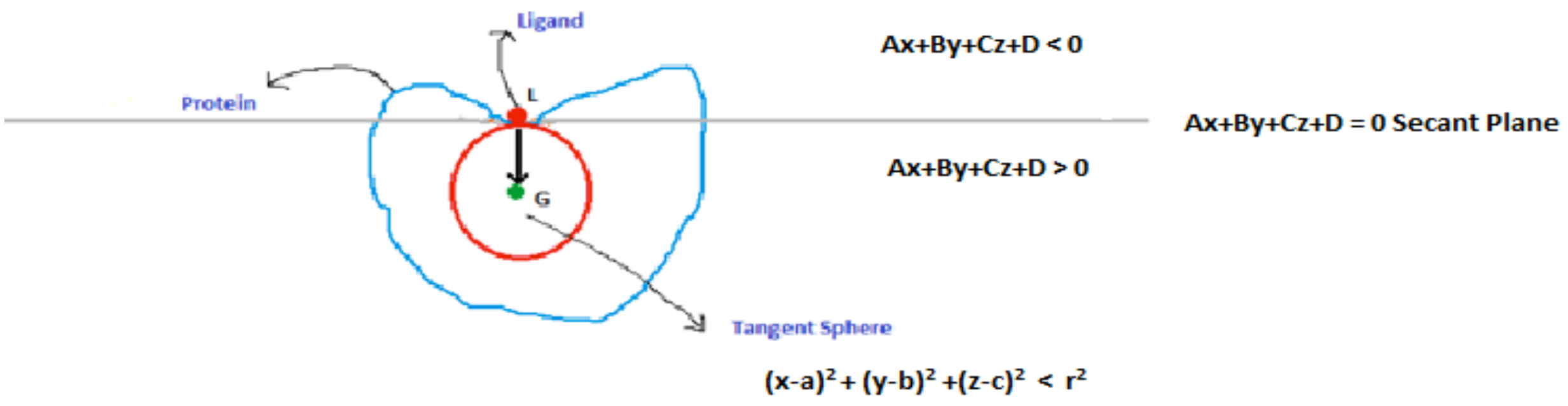

**Figure 4: Equations representing SPM and TSM**

The equations and relations have been devised on the condition that a vector with initial point at the local centroid (as in figure 4) and final point at the global centroid (as in figure 4) is normal to the secant plane at L. The equation of the tangent sphere is given by $\mathbf{(x\text{-}a)^2 + (y\text{-}b)^2 + (z\text{-}c)^2 = r^2}$ and the points that lie inside the tangent sphere is given by $\mathbf{(x\text{-}a)^2 + (y\text{-}b)^2 + (z\text{-}c)^2 < r^2}$ where **r** is the distance between the local centroid and the global centroid and (a, b, c) are the coordinates of the center of the protein.

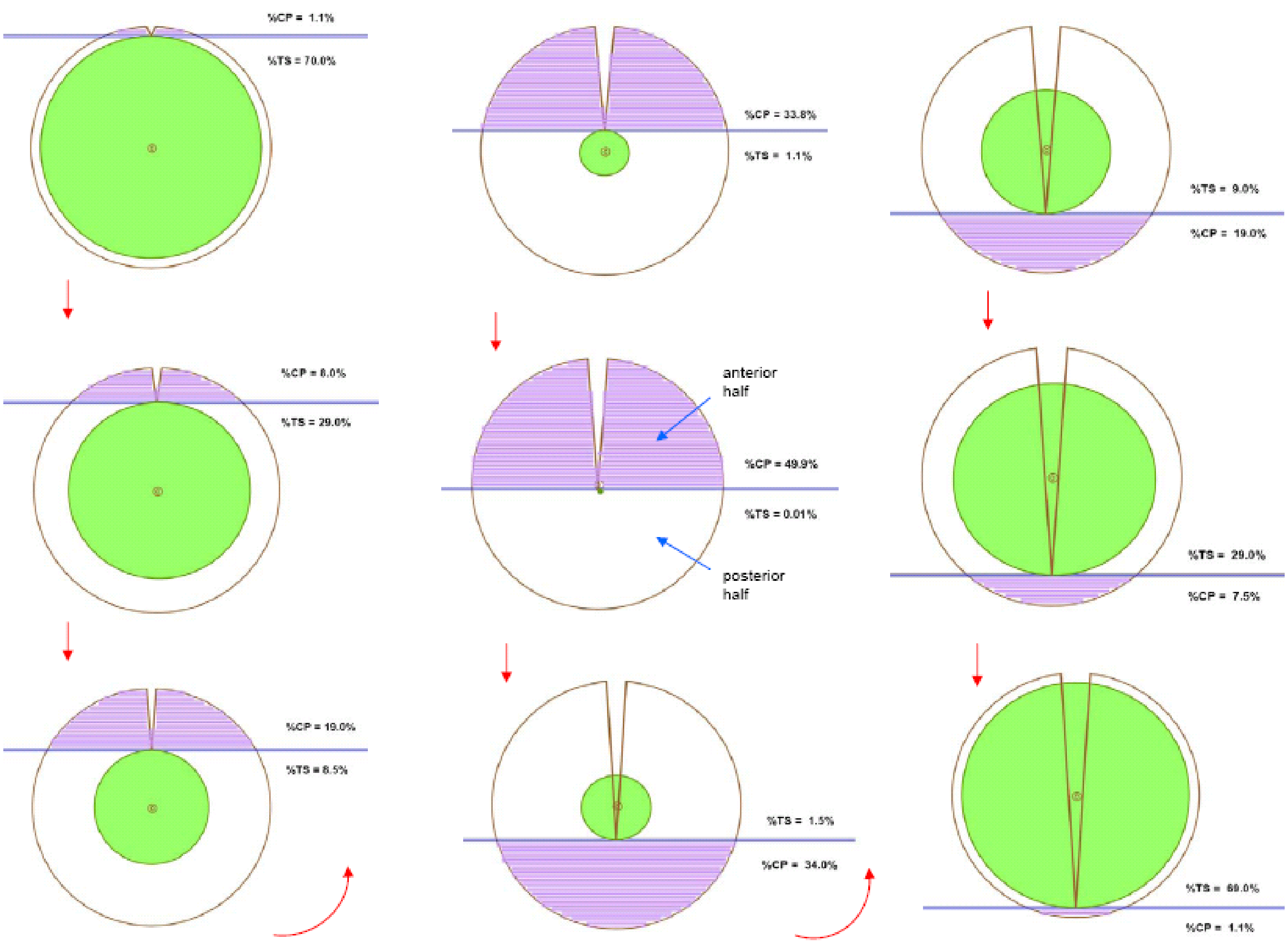

**Figure 5: Depiction of SPi and TSi**

Figure 5 demonstrates the values of SPi and TSi in steps. In the first sphere the burial site depth is minimum, which means SPi is minimum and TSi is maximum. SPi increases gradually from first to fifth sphere whereas TSi decreases simultaneously. In the fifth sphere when the SPi reaches maximum and touches the protein centroid or global centroid the entire process reverses. From sixth to ninth sphere SPi decreases gradually and TSi increases.

## 1.3 Sub methods

There are two sub methods that calculate the binding site depths. We have done all these methods and we call them as below.

(1) **Ligand sub method**: A Protein PDB file and the ligand attached to that particular protein is needed to calculate the TSi and SPi for this method.

(2) **Residue sub method**: This is the more general sub method as it enables SPi and TSi calculation even in the absence of the ligand, as long as the residue knows which amino acids bind the ligand.

## 1.4 PDB

One hundred globular proteins were selected from the PDB, a database consisting of seventy six thousand of protein 3D structures and among these the globular or spherical human proteins are selected by using advanced search parameters such as number of entities and number of models. All the proteins we considered for our research were human proteins as they can be used in for their potential uses in medicine such as drug development research. Each protein might have one or more ligands.

**Visualization Tools**: The protein shape is determined by using the J-Mol visualization tool. Proteins that are roughly spherical or globular were selected.

# Chapter 4

## *1. Web server Implementation and Results*

We developed a web server for the Ligand Burial Site Depth Determination using HTML, Java Script and PHP so that users can upload PDB file of a protein. The server outputs the SPi and TSi (see next sections). The web server is hosted at tortellini.bioinformatics.rit.edu/sxc6274/thesis1.php

There are three file upload options displayed on the web server as in Figure 6. There are two modes in the web server

(1) Ligand Mode: A user can upload both protein PDB file and a ligand PDB file and select the ligand sub method radio button.

(2) Residue Mode: A user can upload both protein PDB file and a residue PDB file (LBS PDB file), and select the residue sub method radio button. The LBS PDB file contains coordinates of one or more amino acids known to be in the LBS.

The minimum number of Ligand or Residue atoms to be uploaded is four (See last sections).All the PDB files to be uploaded in web server should be without header and footer information and it should consist of only the ATOM or HETATM records.

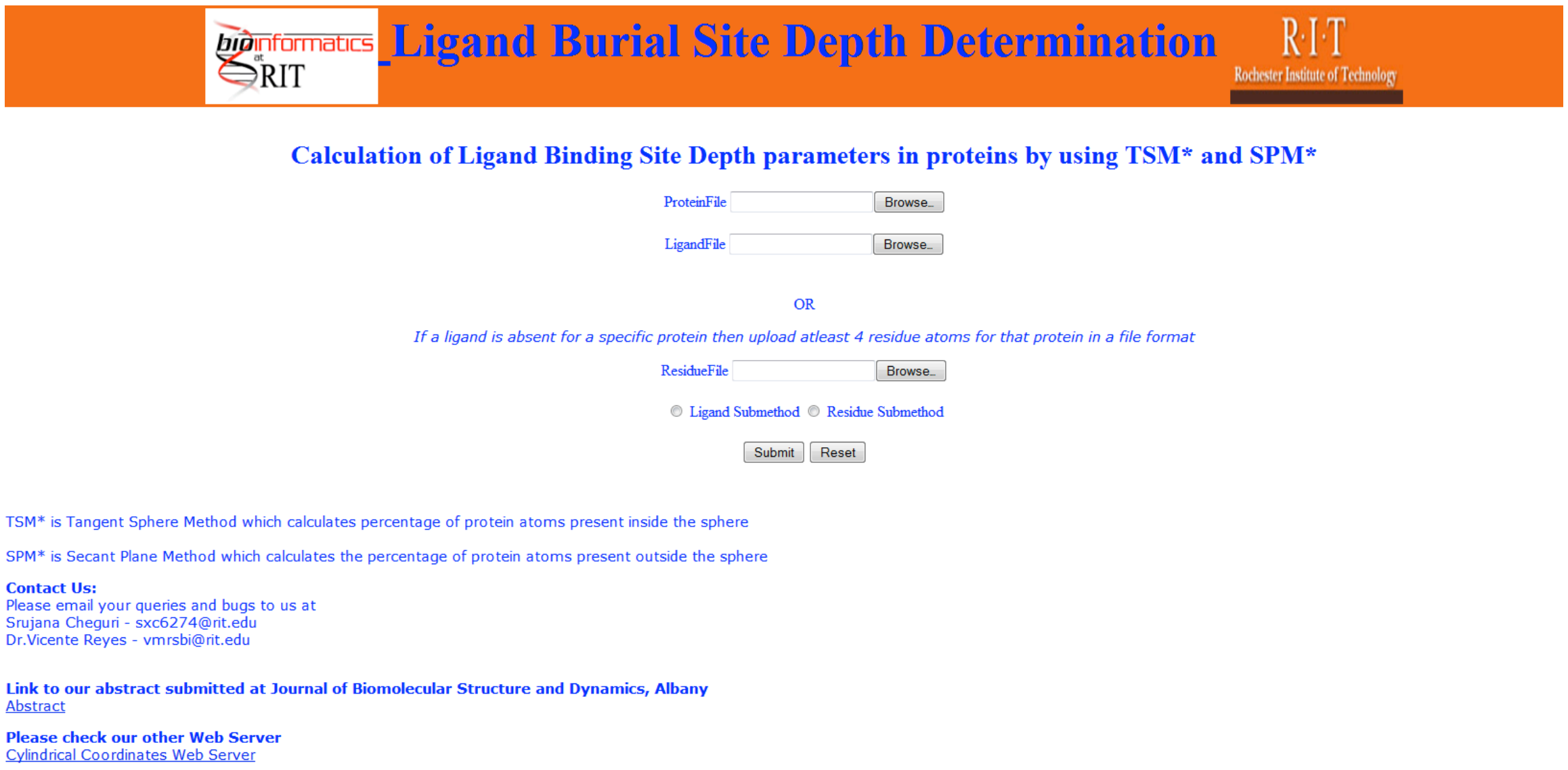

**Figure 6: Home page of "Ligand Burial Site Depth Determination" web server**

## 1.1 Ligand Sub Method

The Ligand sub method is used for protein structures with bound ligands. The PDB coordinates of the protein and the ligand are separated into two files and these two files are the two inputs for this sub method. Shell scripts are written by combining the programs required to calculate SPi and TSi. Two scripts are written separately for this method one each for calculating TSi and SPi. The TSi script comprises of three programs that determine center of mass, radius and TSi for a specific protein-ligand complex. The SPi script consists of three programs that determine Centre of Mass, Secant Plane Coefficients and Secant Plane Index for a protein-ligand complex. The Secant Plane and Tangent sphere indices are displayed on the results web page as in Figure 7.

## 1.2 Residue Sub Method

The residue Sub Method allows calculation of the LBS in the absence of the ligand (i.e., structures that don't contain bound ligand). In this case the user needs to know before hand, which residues in the protein is part of the LBS. PDB coordinates of the protein file and residue file are required for this sub method. Scripts are similar to those of Ligand Sub method, except a residue file is used instead of the ligand file. The SPi and TSi indices will be similarly displayed on the results web page as in Figure 7 upon execution of the scripts.

## 1.3 Results

SPM and TSM methods were applied on about 60 globular protein-ligand complexes. Some of the protein-ligand complexes used for the research are listed in Table1. The depth measurements from the SP and TS methods correlate closely with the ligand pocket size measurements done by visual inspection in the work by Laskowski, et al., (1996; Reyes, V.M. and Cheguri, S.R., manuscript in preparation). This is strong evidence that the SP and TS methods work as they were meant to.

| Protein | ligand |
|---|---|
| 1CMY | HEM |
| 1COH | COH |
| 1DRF | SO4 |
| 1FDH | HEM |
| 1HBS | HEM |
| 1HCO | HEM |
| 1HHO | PO4 |
| 1NIH | HNI |
| 1RNE | NGA |
| 1THB | IHP |
| 2HCO | HEM |
| 2HHB | HEM |
| 2HHM | SO4 |
| 2LOV | CA |
| 2WMB | MG |
| 2WR6 | ODT |
| 2XCG | FA8 |
| 2XDK | XDK |
| 2XDL | 2DL |
| 2XDT | EDO |
| 2XFN | FAD |
| 2XFO | FA8 |
| 2XFP | ISN |
| 2XFQ | RAS |
| 2XHR | COP |
| 2XP2 | VGH |
| 2XRE | CO |
| 2XRF | URA |
| 3A5N | ATP |
| 3A7E | SAM |
| 3AGM | A67 |
| 3BSZ | RTL |
| 3F7H | LI |
| 3GPD | SO4 |
| 3GT9 | ZN |
| 3HFW | ADP |
| 3HHB | HEM |
| 3I25 | MV7 |
| 3ID8 | MRK |
| 3ILG | SR |
| 3INC | NI |
| 3IR0 | CU |
| 3K5U | PFQ |

**Table 1: List of Protein-ligand complexes**

**Web Server Results:** If the user uploads PDB format text files and selects the respective sub method the files are stored in the web server with distinct names under a temporary directory. Then the appropriate shell script is called and applied on the saved files stored in the temporary web directory and the SPi and TSi are calculated and displayed on the web page. The results page should look similar to the figure 7.

Results Page Screen shot:

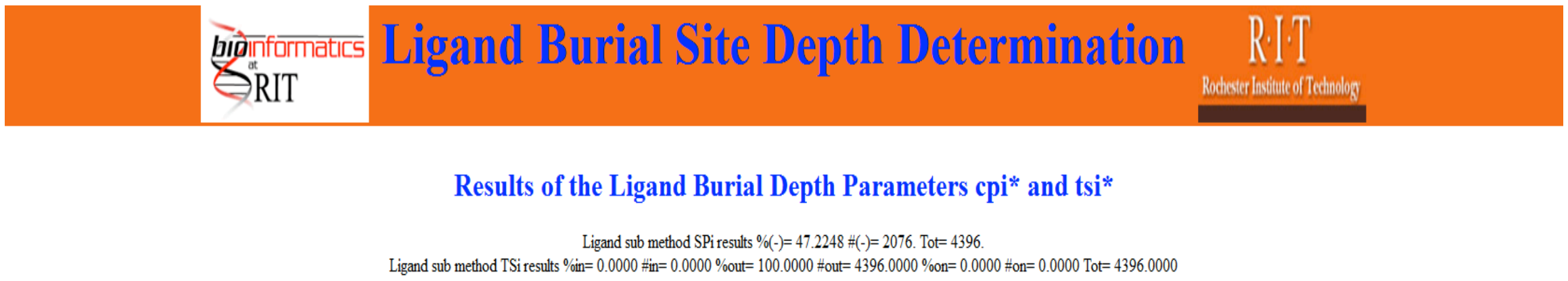

**Figure 7: Result page of "Ligand Burial Depth Determination"**

## 1.4 Validation

Two types of validation are performed on the web server. They are

**Server Side Validation:** This validation is performed at the server side (back end).

This includes checking the format of the uploaded files; it prompts the user if it is not in a right format. It also checks for the size of the uploaded file. The following error messages will be displayed on the results page if there are problems with file uploads:

a) If the user uploads only a protein file and hits the submit button the error message is: "`Not uploaded right type of files`"

b) If the respective sub method is not selected then user will be redirected to a blank result page.

c) If non-text files and files greater than 1MB are uploaded the error message is: "`Error: Only text files and below the size of 100,000 KB are accepted`"

d) If the file entered is a text file and not in PDB format the error message is: "`Not in PDB Format`"

The Reset button sets the server back to original.

**Client Side Validation:** It is performed at the client side (front end). It includes generation of an alert button, when the user hits the submit button before uploading the files. An alert button is displayed as shown in the Figure 8.

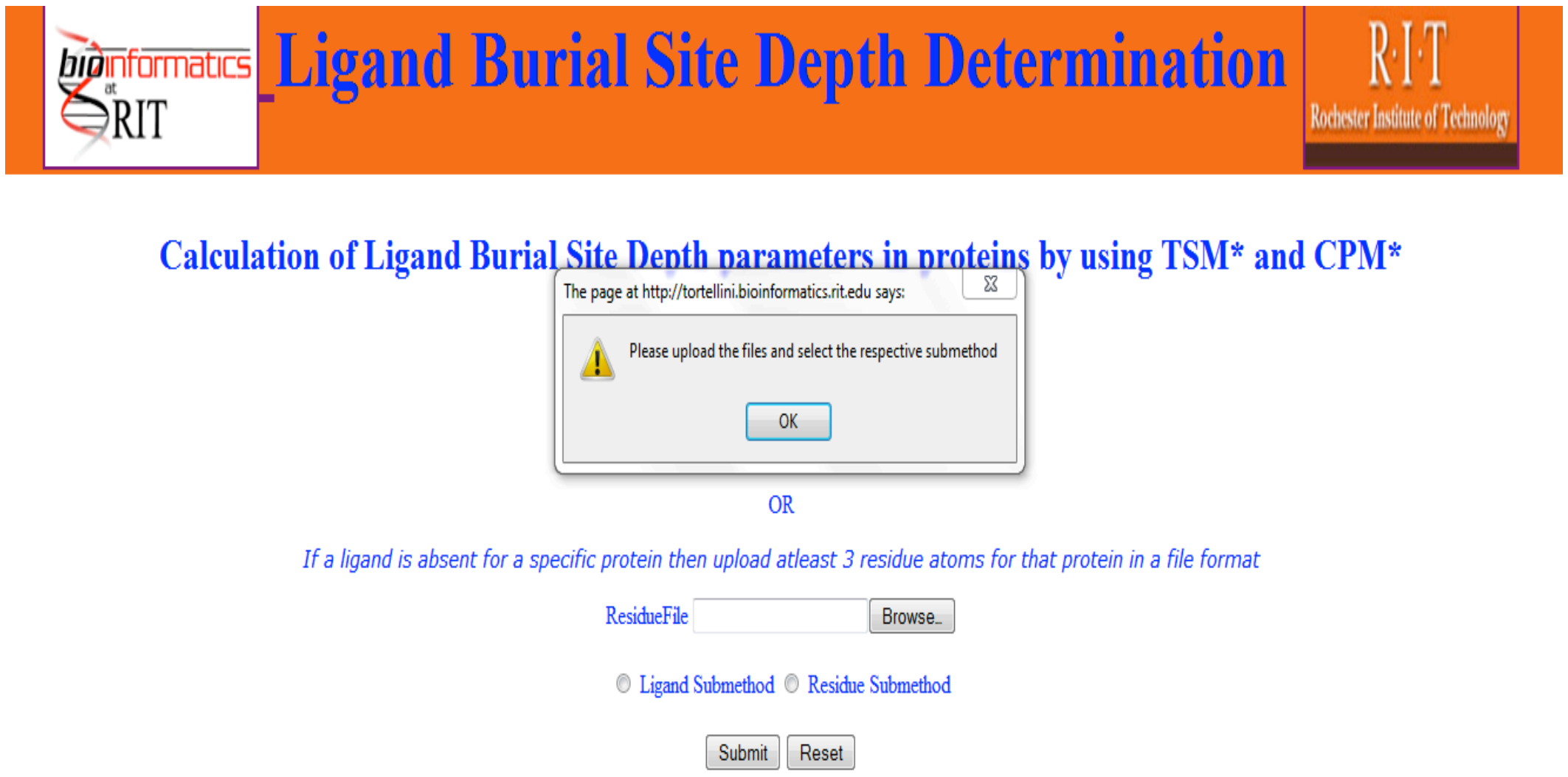

**Figure 8: Alert button in "Ligand Burial Depth Determination" webserver**

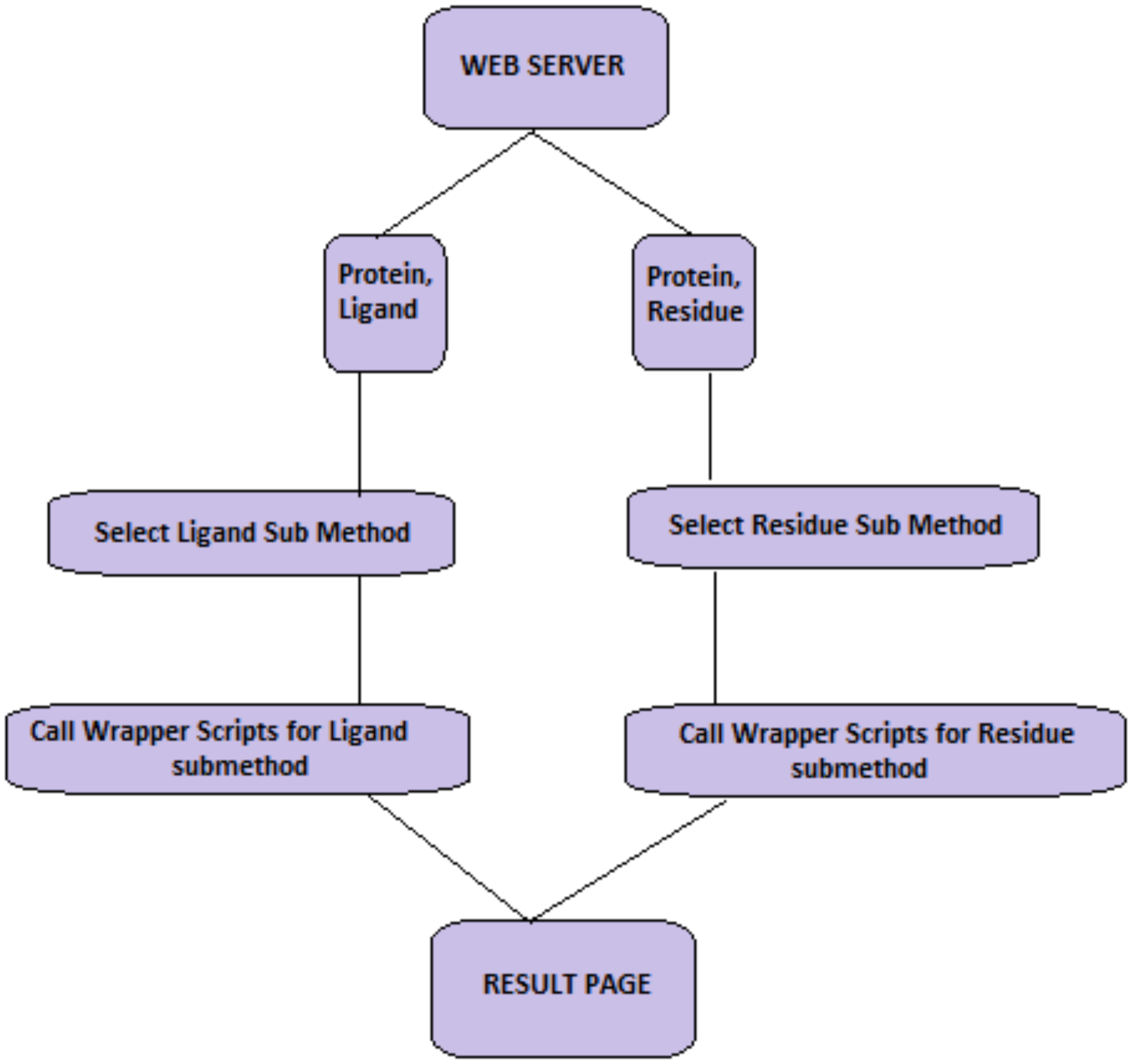

**Figure 9: Flow chart of the steps in "Ligand Burial Depth Determination" web server**

Figure 9 describes the sequence of steps carried out at the back-end of the web server once the files are uploaded.

# Chapter 5

## *1. Challenges and Conclusion*

### *Problem with the PDB format*

We initially observed that the results obtained in UNIX command line method were different compared to the results from the web server (GUI). The problem here is the inconsistent format of the PDB files. The inconsistencies include:

a) Some files differ in the effective number of columns due to merging of two neighboring columns, eliminating the space between them.

b) Some files have misaligned atoms.

The problem arises when we perform a sort step, where the sorting is done on a specified column number. We partially solved this problem by executing a pre-processing step before calling any other program in the script. The pre-processing program written in FORTRAN, rearranges the first few columns in the PDB file so that the number of effective columns in the uploaded files are consistent.

## References:

1. Ask the CSS Guy. (2011). Form field hints with CSS and JavaScript. Retrieved March 20 2011 from http://www.askthecssguy.com/2007/03/form_field_hints_with_css_and.html
2. Berman, H. M., Westbrook, J., Feng, Z., Gilliland, G., Bhat, T. N., Weissig, H., Bourne, P. E. (2000). The protein data bank. *Nucleic Acids Research, 28*(1), 235-242.
3. Binding Site. (2011). Binding site of dihydrofolate reductase (3dfr). Retrieved July 11 2011, from http://www.biochem.ucl.ac.uk/~roman/surfnet/examples/3dfr_site.html
4. Brady, G. P.,Jr, & Stouten, P. F. (2000). Fast prediction and visualization of protein binding pockets with PASS. *Journal of Computer-Aided Molecular Design, 14*(4), 383-401.
5. Coleman, R. G., & Sharp, K. A. (2006). Travel depth, a new shape descriptor for macromolecules: Application to ligand binding. *Journal of Molecular Biology, 362*(3), 441-458. doi:10.1016/j.jmb.2006.07.022
6. Hendlich, M., Rippmann, F., & Barnickel, G. (1997). LIGSITE: Automatic and efficient detection of potential small molecule-binding sites in proteins. *Journal of Molecular Graphics & Modelling, 15*(6), 359-63, 389.
7. Introductory Biology, Cornell University. (2011) Fibrous vs Globular Proteins.Retrieved July 15 2011 from http://www.biog11051106.org/demos/105/unit1/fibrous_v_glob.html
8. Jmol. (2011). Jmol: an open-source Java viewer for chemical structures in 3D. Retrieved July 15, 2011 from http://www.jmol.org
9. Laskowski, R. A., Luscombe, N. M., Swindells, M. B., & Thornton, J. M. (1996). Protein clefts in molecular recognition and function. *Protein Science : A Publication of the Protein Society, 5*(12), 2438-2452. doi:10.1002/pro.5560051206
10. Laurie, A. T., & Jackson, R. M. (2005). Q-SiteFinder: An energy-based method for the prediction of protein-ligand binding sites. *Bioinformatics (Oxford, England), 21*(9), 1908-1916. doi:10.1093/bioinformatics/bti315
11. PHP. (2011). PHP Tutorial. Retrieved July 15 2011 from http://www.w3schools.com/php/

12. Pocket-Finder Pocket Detection. (2011). Detection of Pockets using Pocket Finder Retrieved July 15 2011 from http://www.modelling.leeds.ac.uk/pocketfinder/
13. Shen, S., & Tuszynski, A. (2008). *Theory and Mathematical methods in Bioinformatics*. Berling: Germany.
14. Sheth, V.N., (2009). Visualization of Protein 3D Structures in Reduced Representation with Simultaneous Display of Intra and Inter-molecular Interactions (Master's Thesis). Retrieved March 15 2011from http://rc.rit.edu/docs/sheth-thesis-10.09.pdf
15. thesitewizard.com. (2011). Form Input Validation JavaScript. Retrieved March 20 2011 from http://www.thesitewizard.com/archive/validation.shtml
16. Voet D, Voet J.G. *Biochemistry*. New York: John Wiley and Sons, 1990. Print.
17. Weisel, M., Proschak, E., & Schneider, G. (2007). PocketPicker: Analysis of ligand binding-sites with shape descriptors. *Chemistry Central Journal, 1*, 7. doi:10.1186/1752-153X-1-7

# Appendix

## PART-1

### Script for Ligand Sub Method

cp residue file1; ./pre_process.x; mv file2 filei;./find_CM.x; mv fileo filea; cp protein file1; ./pre_process.x; cp file2 filei; ./find_CM.x;
mv fileo fileb; ./find_CP_coeffs.x; mv fileo filea; cp file2 fileb; ./CPM_NegSid.x;

### Script for Residue Sub Method

cp protein file1; ./pre_process.x; mv file2 filei; ./find_CM.x; mv fileo filea; cp residue file1; ./pre_process.x; mv file2 filei; ./find_CM.x;
mv fileo fileb; ./find_len_2pts.x; cp fileo fileb; cp protein file1; ./pre_process.x; mv file2 filec; ./tangent_sphere_method.x;

### Front end code

```
<html>

<head>

<!-- Reference Form Input Validation
(http://www.thesitewizard.com/archive/validation.shtml)-->

     <script type="text/javascript" language="javascript">

     function validate_form (form )

     {

    if( form.protein.value == "")

          {

                    alert("Please upload the files and select the respective
submethod");

                    form.protein.focus();

                    return false;

          }

          return true;

     }

     function trim(str)

     {
```

```
            return str.replace(/^\s+|\s+$/g,'');
                        }
</script>
<!-- Reference Form Input Validation Ends here-->
<style type="text/css">
h1
 {
            background-color: #F36E21;
            border-style: solid;
            border-width: 3px;
            border-left-width:5px;
            border-right-width:5px;
            border-color: #F36E21;
            margin-top: -8.5px;
            margin-right: -5px;
            margin-left: -5px;
            <!--margin: 0.5em;-->
            padding:4em;
            font-size:50px;
            }
h2
{
font-family:font-family:Berlin Sans FB;
font-size:27px;
}
p
{
```

```
font-family:Verdana;

font-size:15px;

}

</style>

</head>

<body>

<center> <font color="blue">

<h1> <a href="http://www.bioinformatics.rit.edu">

<img align=center src="bioinfologo.gif" alt="Bioinformatics at RIT" height="100"
width="180"> </img> </a>

Ligand Burial Site Depth Determination <a href="http://www.rit.edu">

<img align=center src="rit.jpg" alt="Rochester Institute of Technology" height="100"
width="180"> </img> </a></h1>

<h2>Calculation of Ligand Burial Site Depth parameters in proteins by using TSM* and
CPM*  </h2>

<form action="result1.php" name="file_form" method="post" enctype="multipart/form-
data" onSubmit="return validate_form (this );"  >

ProteinFile <input type="file" name="protein" input type="hidden"
name="MAX_FILE_SIZE" value="1000000" />

<br /> <br />

LigandFile <input type="file" name="ligand" input type="hidden"
name="MAX_FILE_SIZE" value="500000" />

 <br /> <br /> <br />

 OR

 <br />

 <p><i>If a ligand is absent for a specific protein then upload atleast 3 residue atoms for
that protein in a file format</i> </p>

ResidueFile <input type="file" name="residue" input type="hidden"
name="MAX_FILE_SIZE" value="500000" />

 <br /> <br />
```

```
<input type="radio" name="submethod" value="Ligand Submethod"/> Ligand Submethod

<input type="radio" name="submethod" value="Residue Submethod" /> Residue Submethod

<br /> <br />

<input type="submit" value="Submit"  />

<input type="reset" value="Reset" />

</form>

</center>

<br />

<left>

<p> TSM* is Tangent Sphere Method which calculates percentage of protein atoms present inside the sphere <br /> <br />

CPM* is Secant Plane Method which calculates the percentage of protein atoms present outside the sphere </p>

<p> <b> Contact Us: </b> <br />Please email your queries and bugs to us at <br />

Srujana Cheguri - sxc6274@rit.edu <br />

Dr.Vicente Reyes - vmrsbi@rit.edu

<br /> <br />

<p><b>Please check our other Web Server <a href="http://tortellini.bioinformatics.rit.edu/sxc6274/thesis2.php"></b><br />

Cylindrical Coordinates Web Server</a></p>  </p>

<br />

</left>

</font>

</body>

</html>
```

**Back end code**

```
<html>
        <head>
                <title>
                        Results Page
                </title>
<style type="text/css">
h1
  {
        background-color: #F36E21;
        border-style: solid;
        border-width: 3px;
        border-left-width:5px;
        border-right-width:5px;
        border-color: #F36E21;
        margin-top: -8.5px;
        margin-right: -5px;
        margin-left: -5px;
        <!--margin: 0.5em;-->
        padding:4em;
        font-size:50px;
        }
h2
{
font-family:font-family:Berlin Sans FB;
font-size:27px;
}
```

```
p
{
font-family:Verdana;
font-size:15px;
}
</style>
</head>
<body >
<center>
<font color="blue">
<h1> <a href="http://www.bioinformatics.rit.edu">
<img align=center src="bioinfologo.gif" alt="Bioinformatics at RIT" height="100"
width="180"> </img> </a>
Ligand Burial Site Depth Determination <a href="http://www.rit.edu">
<img align=center src="rit.jpg" alt="Rochester Institute of Technology" height="100"
width="180"> </img> </a></h1>
<h2>Results of the Ligand Burial Depth Parameters SPi* and tsi* </h2>
<center>
 </body>
</font>
</html>
<?php
#Reporting the simple errors
// Report simple running errors
ini_set('error_reporting', E_ALL ^ E_NOTICE);
// Set the display_errors directive to OFF
ini_set('display_errors', 0);
```

```
// Log errors to the web server's error log
ini_set('log_errors', 1);
// Destinations
define("ADMIN_EMAIL", "sxc6274@rit.edu");
define("LOG_FILE", "/home/sxc6274/public_html/error1.log");
// Destination types
 define("DEST_EMAIL", "1");
  define("DEST_LOGFILE", "3");
/* Examples */
// Send an e-mail to the administrator
#error_log("Fix me!", DEST_EMAIL, ADMIN_EMAIL);
// Write the error to our log file
error_log("Error", DEST_LOGFILE, LOG_FILE);
function my_error_handler($errno, $errstr, $errfile, $errline)
        {
        switch ($errno) {
                case E_USER_ERROR:
                        // Send an e-mail to the administrator
                        error_log("Error: $errstr \n Fatal error on line $errline in file
$errfile \n", DEST_EMAIL, ADMIN_EMAIL);


                        // Write the error to our log file
                        error_log("Error: $errstr \n Fatal error on line $errline in file
$errfile \n", DEST_LOGFILE, LOG_FILE);
                        break;


                case E_USER_WARNING:
```

```
                        // Write the error to our log file

                        error_log("Warning: $errstr \n in $errfile on line $errline \n",
DEST_LOGFILE, LOG_FILE);

                        break;


                case E_USER_NOTICE:

                        // Write the error to our log file

                        error_log("Notice: $errstr \n in $errfile on line $errline \n",
DEST_LOGFILE, LOG_FILE);

                        break;


                default:

                        // Write the error to our log file

                        error_log("Unknown error [#$errno]: $errstr \n in $errfile on line
$errline \n", DEST_LOGFILE, LOG_FILE);

                break;

        }


        // Don't execute PHP's internal error handler

        return TRUE;

        }// Use set_error_handler() to tell PHP to use our method

$old_error_handler = set_error_handler("my_error_handler");

#include('/home/sxc6274/config.php');

#http://myphpform.com/validating-forms.php

$file1 = input_val($_POST["protein"]);

#chmod("$file1",0777);

$file2 = input_val($_POST["ligand"]);

#chmod("$file2",0777);
```

```
$submethod = input_val($_POST["submethod"]);
$file3 = input_val($_POST["residue"]);
#chmod("$file3",0777);

function input_val($data)
	{
	$data = trim($data);
	$data = stripslashes($data);
	$data = htmlspecialchars($data);
	return $data;
	}
$filename1 = basename($_FILES["protein"]["name"]);

#echo $filename1;
#chmod("$filename1",0777);
$filename2=basename($_FILES["ligand"]["name"]);
#echo $filename2;
#chmod("$filename2",0777);
$filename3=basename($_FILES["residue"]["name"]);
#echo $filename3;
#chmod("$filename3",0777);
$findkey1='ATOM';
$findkey2='HETATM';
$ext1=substr($filename1,strpos($filename1,'.')+1);
$ext2=substr($filename2,strpos($filename2,'.')+1);
$ext3=substr($filename3,strpos($filename3,'.')+1);
$target_path = "/home/sxc6274/public_html/uploads/";
```

```
$newname1 = $target_path.$filename1;

#echo "$newname1 <br>\n";

#chmod("$newname1",0777);

$newname2 = $target_path.$filename2;

#echo "$newname2 <br> \n";

#chmod("$newname2",0777);

$newname3 = $target_path.$filename3;

#echo "$newname3 <br> \n";

#chmod("$newname3",0777);

$dirname=substr($filename1,0,4);

#echo "$dirname <br> \n";

#$dirname2=substr($filename2,0);

$check1='($ext1=="txt") && ($_FILES["protein"]["size"] < 1000000)';

#echo "$check1 <br> \n";

$check2='($ext2=="txt") && ($_FILES["ligand"]["size"] < 1000000)';

#echo "$check2 <br> \n";

$check3= '($ext3=="txt") && ($_FILES["residue"]["size"] < 1000000)';

#echo "$check3 <br> \n";

 if (($check1 && $check2) || ($check1 && $check3))

  {  #echo "yes first if loop";

  if((is_uploaded_file($_FILES['protein']['tmp_name'])) &&
((is_uploaded_file($_FILES['ligand']['tmp_name']))||(is_uploaded_file($_FILES['residue'
]['tmp_name']))))

            {

#echo "second ";

$handle1 =fopen($_FILES['protein']['tmp_name'], "r");

                    $handle2 =fopen($_FILES['ligand']['tmp_name'], "r");
```

```
                    $handle3 =fopen($_FILES['residue']['tmp_name'], "r");
                    $line1=fgets($handle1);
                    #echo "$line1 line1<br> \n";
                    $line2=fgets($handle2);
                    #echo "$line2 line2<br> \n";
          $line3=fgets($handle3);
            #echo "$line3 line3<br> \n";
            $pos1=strpos($line1,'ATOM');
                    #echo "$pos1 for pos1 <br> \n";
                    $pos2=strpos($line2,'HETATM');
                    #echo "$pos2 for pos2 <br> \n";
                    $pos3=strpos($line3, 'ATOM');
                    #echo "$pos3 for pos3 <br> \n";
                    if(($pos1===FALSE)&&(($pos2===FALSE)||($pos3===FALSE)))
                    {
            echo "Not a PDB file <br> \n";
                            exit(1);
                    }
     else
                    {
            while((!feof($handle1)) && (!feof($handle2))||(!feof($handle3)))
                            {
                            $line1=fgets($handle1);
                                    $line2=fgets($handle2);
                                    $line3=fgets($handle3);
     if(($pos1 == 0)&&(($pos2 == 0)||($pos3 == 0)))
                                    {
```

```
$nf1=preg_split("/[\s,]+/",$line1);
                                        $x=sizeof($nf1);
                                        #echo "$x for nf1<br> \n";
                                        $nf2=preg_split("/[\s,]+/",$line2);
                                        $y=sizeof($nf2);
                                        #echo "$y for nf2 <br> \n";
                                        $nf3=preg_split("/[\s,]+/",$line3);
                                        $z=sizeof($nf3);
                                        #echo "$z for nf3 <br> \n";
            $cond1=(($x==13)||($x==12));
            $cond2=(($y==13)||($y==12));
            $cond3=(($z==13)||($z==12));
          if (($cond1 && $cond2)||($cond1 && $cond3))
                                        {
                                             break;
                                             #echo"checking if or condition";
                                        }
                                        else
                                        {
                                             echo "Not in PDB Format <br> \n";
                                             exit(1);
                                        }
                                   }
                              }
                         }
                    }
```

```
            else
            {
                        echo "Not uploaded right type of files <br> \n";
                        exit(1);
            }

 $expr1=(!file_exists($newname1));
 #echo "$expr1 for expr1 <br> \n";
 $expr2=(!file_exists($newname2));
 #echo "$expr2 for expr2 <br> \n";
 $expr3=(!file_exists($newname3));
# echo "$expr3 for expr3 <br> \n";
$exprval1=($expr1 && $expr2);
#echo "$exprval1 for expr1 <br> \n";
$exprval2=($expr1 && $expr3);
#echo "$exprval2 for expr2 <br> \n";
            if(($expr1 && $expr2) || ($expr1 && $expr3))
            {
            @system("rm -r /home/sxc6274/public_html/uploads/$dirname");
            #echo "entered file_exists if loop <br> \n";
$rval1= move_uploaded_file($_FILES['protein']['tmp_name'],$newname1);
                        $rval2 =
move_uploaded_file($_FILES['ligand']['tmp_name'],$newname2);
                        $rval3 =
move_uploaded_file($_FILES['residue']['tmp_name'],$newname3);
                        $rval = $rval1 && $rval2;
                        $rvalx = $rval1 && $rval3;
```

```
if(($rval)||($rvalx))

{

#echo "succesfully entered the loop of move uploaded file";

@system("mkdir /home/sxc6274/public_html/uploads/$dirname");

@system("mv $newname1
/home/sxc6274/public_html/uploads/$dirname/protein");

@system("mv $newname2
/home/sxc6274/public_html/uploads/$dirname/ligand");

@system("mv $newname3
/home/sxc6274/public_html/uploads/$dirname/residue");

#chdir ("/home/sxc6274/public_html/files/$dirname");

#chdir ("/home/sxc6274/public_html/files");

@system("cp /home/sxc6274/public_html/uploads/pre_process.f
/home/sxc6274/public_html/uploads/$dirname");

@system("cp /home/sxc6274/public_html/uploads/pre_process.x
/home/sxc6274/public_html/uploads/$dirname");
@system("cp /home/sxc6274/public_html/uploads/CPM_NegSid.f
/home/sxc6274/public_html/uploads/$dirname");

@system("cp
/home/sxc6274/public_html/uploads/CPM_NegSid.x
/home/sxc6274/public_html/uploads/$dirname");

@system("cp
/home/sxc6274/public_html/uploads/find_CM.f
/home/sxc6274/public_html/uploads/$dirname");

@system("cp
/home/sxc6274/public_html/uploads/find_CM.x
/home/sxc6274/public_html/uploads/$dirname");

@system("cp
/home/sxc6274/public_html/uploads/find_CP_coeffs.f
/home/sxc6274/public_html/uploads/$dirname");

@system("cp
/home/sxc6274/public_html/uploads/find_CP_coeffs.x
/home/sxc6274/public_html/uploads/$dirname");
```

```
@system("cp
/home/sxc6274/public_html/uploads/find_len_2pts.f
/home/sxc6274/public_html/uploads/$dirname");

@system("cp
/home/sxc6274/public_html/uploads/find_len_2pts.x
/home/sxc6274/public_html/uploads/$dirname");

@system("cp
/home/sxc6274/public_html/tangent_sphere_method.f
/home/sxc6274/public_html/uploads/$dirname");

@system("cp
/home/sxc6274/public_html/tangent_sphere_method.x
/home/sxc6274/public_html/uploads/$dirname");

if($submethod=='Ligand Submethod')

{

@system("cp /home/sxc6274/public_html/uploads/ligand_SPi.sh
/home/sxc6274/public_html/uploads/$dirname");


#chmod("/home/sxc6274/public_html/uploads/$dirname",0777);

chdir ("/home/sxc6274/public_html/uploads/$dirname");

#system('pwd');

#echo "entered ligand submethod <br>\n";

@system("./ligand_SPi.sh");

$out1=file_get_contents('filez');

echo "Ligand sub method SPi results $out1 <br>\n";

@system("cp
/home/sxc6274/public_html/uploads/ligand_tsi.sh
/home/sxc6274/public_html/uploads/$dirname");

#system('pwd');

@system("./ligand_tsi.sh");

$out2=file_get_contents('fileg');

echo "Ligand sub method TSi results $out2 <br>\n";

}
```

```
elseif($submethod=='Residue Submethod')
{
#chmod("/home/sxc6274/public_html/uploads/$dirname",0777);
chdir ("/home/sxc6274/public_html/uploads/$dirname");
@system("cp /home/sxc6274/public_html/uploads/residue_SPi.sh /home/sxc6274/public_html/uploads/$dirname");
#chmod("/home/sxc6274/public_html/uploads/residue_SPi.sh", 0777);
@system("cp /home/sxc6274/public_html/uploads/residue_tsi.sh /home/sxc6274/public_html/uploads/$dirname");
#chmod("/home/sxc6274/public_html/uploads/residue_tsi.sh", 0777);
@system("./residue_SPi.sh");
$out3=file_get_contents('filez');
echo "Residue sub method SPi results $out3 <br>\n";
@system("./residue_tsi.sh");
$out4=file_get_contents('fileg');
echo "Residue sub method TSi results $out4 <br>\n";
}


}
}
else
{
echo "Error: File ".$_FILES["protein"]["tmp_name"] ." and ".$_FILES["ligand"]["tmp_name"] ." already exists";
}
}
else
```

```
{

echo "Error: Only text files and below the size of 100,000 KB are accepted";

}

?>
```

**PART 2**

**Wrapper Script**

```
cp tip filea; cp protein fileb; ./translate.x;
cp fileo filei; ./Zrotation.x;
cp fileo filei; ./cart2cyl_v2_deg.x;
```

**Front end Code**

```
<html>
<head>
<style type="text/css">
h1
{
   background-color: #513127;
   border-style: double;
   border-width: 3px;
   border-left-width:5px;
   border-right-width:5px;
   border-color: #E67451;
   margin-top: -8.5px;
   margin-right: -5px;
   margin-left: -5px;
   <!--margin: 0.5em;-->
   padding:4em;
   font-size:50px;
   }

h2
{
font-size:34px;
}
p
{
```